\documentstyle[12pt,epsfig]{article}

\newcommand{\beq}{\begin{equation}}
\newcommand{\eeq}{\end{equation}}

\begin{document}

\begin{center}
\vskip 1.75cm

{\Huge {\bf Fast antibaryon production:}}\\
\vskip 0.2cm
{\Huge {\bf A new collective effect}}\\
\vskip 0.2cm
{\Huge {\bf in Pb--Pb collisions?}}\\
\vskip 0.8cm

E. G. Ferreiro and C. Pajares\\ {\sl Departamento de
F\'{\i}sica de  Part\'{\i}culas,\\
Universidade de Santiago de
Compostela,\\ 15706--Santiago de Compostela, Spain}

\vskip 1.0truecm
{\Large {\bf Abstract}}
\end{center}
\begin{quotation}

Recent experimental data on antibaryon production outside 
nucleon--nucleon kinematical limits
have been obtained in Pb--Pb collisions at SPS energies.
We present a possible explanation for this phenomenon
based on the existence of
collective effects. 
String Fusion Model results are compared
to the
data and predictions are also presented.

\end{quotation}
\vspace{2cm}


\vspace{4.5cm}
\begin{flushleft}
{\bf US--FT/18--97}\\
%{\today}
\end{flushleft}

\newpage

%\noindent{\bf 1. Introduction:}                                               

During the last year, new interesting effects have been shown by the recent
data on
Pb--Au and Pb--Pb collisions at $p_{lab}=158$ GeV/c per nucleon.
Increase of e$^{+}$e$^{-}$ pairs in the mass region of 
0.3 GeV/c$^{2}$ $<$ m $<$ 0.7 GeV/c$^{2}$ \cite{ee}, strangeness enhancement
\cite{str} and a much larger $J/\psi$ suppression than in the case of S--U
collisions \cite{supre} are some of these phenomena. 
Many papers have appeared on the physical origin of these effects, discussing
mainly whether or not Quark Gluon Plasma (QGP) has been formed 
\cite{QGPform}--\cite{Sorge}.

There is another part of the experimental data which, up to now, has not been
paid enough attention. In fact, the NA52 Collaboration \cite{NA52} has detected
antiprotons outside of the kinematical limits of nucleon--nucleon collisions.
The existence of particles above the nucleon--nucleon kinematical limit 
(particles with Feynman $x$ bigger than one) in
Pb--Pb collisions, the well--known cumulative effect 
\cite{Bal74}--\cite{Bra94},
was predicted to appear by the action 
of collective mechanisms \cite{xfey}, such as the string fusion or eventually
the percolation of strings \cite{Perco}. 

In this paper we compare our results with the data. 
First, we check that the String Fusion Model Code
(SFMC, \cite{SFMC}) gives a reasonable description of other
observables, like the multiplicity of negative charged particles or
$\overline{\Lambda}$ production in Pb--Pb collisions at SPS energies. 
In previous papers we have shown that the model describes reasonable well the
hadron--hadron, hadron--nucleus and nucleus--nucleus existing data at similar
energies, in particular the strangeness enhancement data \cite{strange} and the
cumulative effect seen in hadron--nucleus collisions at 
400 GeV/c \cite{xfey}.

%\vspace{0.5cm}
%\noindent{\bf 2. The String Fusion Model Code:}
In many models of hadronic collisions,
like the Dual Parton Model (DPM, \cite{Cap94}) or
equivalently the Quark Gluon String Model (QGSM, \cite{Kai84}) and models
based
 on
them like VENUS \cite{Wer87}, 
colour strings are exchanged between projectile and target.
The number of strings grows with the energy and with the number of nucleons of
participant nuclei.

The String Fusion Model Code  
is a Monte Carlo code based on the 
QGSM that incorporates the possibility of string fusion.
When the density of strings becomes high, the string fields begin to overlap
and eventually individual strings may fuse.
We consider that the strings fuse when their transverse
positions come within a certain interaction area, of the same order as
parton--parton one. For present calculations it has been taken equal to 7.5 mb,
and only fusion of strings in groups of 2 has been included.
This value was previously fixed to describe the $\overline{\Lambda}$ production
in S--S and S--Ag collisions \cite{strange}.
The 
quantum numbers of the fused string are determined by the interacting partons
and
its energy--momentum is the sum of the energy--momenta of the ancestor
strings.
 The
colour charges of the fusing strings ends sum into the  colour charge  of the
resulting string ends according to the $SU(3)$ composition laws.
The new strings break into hadrons according to their higher colour. 
As a result, heavy flavour is produced more efficiently and there is a
reduction of the total multiplicity.

On the other hand, since the energy--momenta of 
the original strings are summed to obtain the energy--momentum of the resulting
string, the fragmentation of the latter can produced some particles outside the
kinematical limits of nucleon--nucleon collisions if the original strings come
from different nucleon-nucleon collisions.

The code does not include other mechanisms like pop corn or cascading in spite
of giving rise to non negligible effects. The reason 
is to distinguish the string fusion effects from the rest.

%\vspace{0.5cm}
%\noindent{\bf 3. Results:}
We have run our code in order to reproduce Pb--Pb central (impact parameter 
$b \leq 2.8$ fm) 
collisions at $p_{lab}=158$ GeV/c per nucleon.

Our results of negative charged particles, $K^0_s$ and 
antilambdas for the string
fusion case compared to the experimental data \cite{NA49,NA49b}
are shown in Fig. 1.

It is seen that the 
SFMC results for $h^-$ are slightly higher than the experimental
ones in the central region. 
A  reasonable agreement is obtained for $K^0_s$ and $\overline{\Lambda}$
distributions, while
for $\Lambda$
our result \cite{SQM97}
is lower than the experimental one, 
 due to the fact that we 
do not include the cascading mechanism
in the code. It is very important for 
$\Lambda$ production due to the processes:

\noindent
$\pi^-$p($\pi^0$n)$\rightarrow$$K^+$$\Sigma^-$, $\overline{K^0}$$\Lambda$
\ \ ;
\ \  $\pi^+$n($\pi^0$p)$\rightarrow$$K^+$$\Sigma^0$, $K^+$$\Lambda$,
                               $\overline{K^0}$$\Sigma^+$ \ \  and \ \
$K^-$p$\rightarrow$$\pi^0$$\Lambda$, $\pi^+$$\Sigma^-$.

%%%%%%%%%%%%%%%%%%%%%%%%%%%%%%%%%%%%%%%%%%%%%%%%%%%%%%%%%%%%
Relative to the cumulative effect 
\cite{Bal74}--\cite{xfey},
in
the string fusion approach it is naturally explained by the increased
longitudinal momentum of the fused strings. String
fusion picks up several partons coming from different nucleon--nucleon
collisions so that particles with higher energy than the initial
nucleon--nucleon one can be obtained.

In the string fusion picture, this can be understood as a process that takes
place in two steps. The first step corresponds to the formation of 4 strings in
2 different nucleon-nucleon collisions, where 2 projectile nucleons and 2
target nucleons are involved. Each nucleon-nucleon collision leads to the
creation of 2 colour strings. The energy of each collision is
divided
into the two formed strings in different fractions, $x$ and $x^{\prime}$. 
So in one collision we will
have 2 strings, whose total energy has to be
$x_1+x_1^{\prime}=1$ (normalized to the total energy of the collision
$\sqrt{s_{NN}}$), and $x_2+x_2^{\prime}=1$ 
for the 2 strings formed in the other 
collision.
The second step arrives when the string fusion mechanism takes place. A fused
string coming from 2 ancestor strings, each of them
created in a
 different nucleon-nucleon collision, is formed.
The energy--momentum of this new string is the sum of the energy--momenta of
the elemental strings.  The cumulative effect ($|x_F|>1$)
can happen when, by adding these energy-momenta, the total fraction of 
energy achieved by
the fused string is
bigger than one, $x_1+x_2 > 1$.

%%%%%%%%%%%%%%%%%5
Based on this idea, we propose the study of the cumulative production as a
trail of the existence of collective effects.

%%%%%%%%%%%%%%%%%%%%%%%%%%%%%%%%%%%
In Fig. 2 our results for the antiproton rapidity
distribution obtained with the SFMC are
compared to the preliminary data
of NA52. A very good agreement is found
between the
SFMC results for the string fusion case and the experimental data.

It is easy to calculate which is the $x_F$  that corresponds to these
antiprotons. These particles, localized at $y=6$ rapidity units and detected at
$p_T=0$,  have a longitudinal momentum $p_z$ 
that can be calculated by using the
well known relation between $y$ and $p_z$:
\beq
y=\frac{1}{2} \ln \frac{E+p_z}{E-p_z} \ \ .
\eeq

The variable  $x_F$ is defined in the center of mass frame. By applying the
usual kinematic relations, it is possible to redefine it 
as a function of the laboratory
frame variables. So we can express  $x_F$ by:
\beq
x_F^{cm}=p_z^{cm}/p_{beam}^{cm}=
-(E_{lab}/m_2)+[(E_{beam}+m_2)p_z^{lab}/(m_2 \ \ p_{beam})] \ \ ,
\eeq
where $E_{lab}$ and $p_z^{lab}$ are the energy and longitudinal
momentum of the particle in the laboratory frame, $m_2$ is the mass of the
target and $E_{beam}$ and $p_{beam}$ are the energy and longitudinal
momentum of the beam in the laboratory frame. As the energy and momentum of the
beam are given per nucleon, then $m_2$ is the nucleon mass.
So the Feynman $x$ corresponding to these antiprotons with $y=6$ will be
$x_F=1.2$, well above the kinematical limits. 

In Fig. 2 we also show the corresponding $x_F^{cm}$ distribution of those
antiprotons. Fermi motion of the nucleons of the nucleus is included in the
code, but the production of antibaryons at this rapidity cannot be explained
without the inclusion of other collective effects. As can be seen in Fig. 2,
the maximum rapidity attained
by the antiprotons when the string fusion mechanism
is not included (so only Fermi motion contributes) is $y=5.5$, that corresponds
to $x_F=0.72$, far away from the kinematical limits.

The preceding results strongly support the idea about the intervention of a
mechanism that permits to add energy coming from different nucleon-nucleon
collisions, as the string fusion does. Other mechanisms of interaction of
strings can also produce very fast particles \cite{Bo}.
%%%%%%%%%%%%%%%

On the other hand, in Fig. 3
our results on baryon production for
 Pb--Pb central collisions at SPS energy are shown.
As we have said above, 
Fermi motion of the nucleons of the nucleus is included in the 
SFMC code. Because
of this, some particles with $x_F>1$ can also be obtained in the no fusion 
case,
but its production is strongly increased by the introduction of the string
fusion.
The NA49 Collaboration is now analyzing data which could test this prediction.
Concerning to the production of mesons with $x_F>1$, in these collisions 
it is possible to obtain
pions with $x_F=1.5$ when the string fusion mechanism
is included, while in the no fusion case the quantity
of mesons found with $x_F$ bigger than one
is negligible.

The results obtained for Pb--Pb central collisions at RHIC energies are similar
to the ones got at SPS energies. For 1000 events, 2015 particles are produced
with $|x_F|>1$, to be compared to 1783 particles found at SPS energies. There
is not a large increase of the cumulative effect from SPS to RHIC energies. 

%%%%%%%%%%%%%%%

Nevertheless, it 
is important to take into account that in the SFMC only fusion of strings
in pairs has been included. 
The string fusion effect increases with the energy and the
atomic number
of 
participant nuclei, 
because the density of strings grows, so the probability for them to fused
becomes higher.
Then for Pb--Pb central collisions at SPS energies or for nucleus--nucleus
collisions at RHIC energies it is necessary to consider the possibility to
fuse the strings not only in pairs but in bigger groups.
In that case, it would be possible to obtained particles with $x_F$ equal to
three or even larger.

Even more, if the density of strings exceeds a critical value that can be
calculate knowing the radius of each string (around
0.2 fm), percolation of strings
becomes possible \cite{Perco}.
The critical density necessary to have percolation is about
$n_c=9$ strings/fm$^2$. Above it, paths of overlapping strings 
(that can be represented as circles in the
transverse space) are formed through the whole collision area. Along
these paths the medium behaves as a colour conductor. 
This critical density is already reached in central Pb--Pb collisions at 158
AGeV/c and in central Ag--Ag collisions at RHIC energies \cite{Perco}.

The region where several strings fuse can be considered as a droplet of a
non--thermalized QGP. Percolation means that these droplets overlap and QGP
domain becomes comparable to nuclear size. 

It is important to take this possibility into account when studying the
cumulative effect. In case of percolation 
the possibility to obtain particles with $x_F$ many times bigger than one would
become no negligible.

More measurements of this effect in the actual Pb--Pb
experiments 
and at RHIC will be welcome
in order to clarify the existence of collective effects 
such as string fusion or eventually percolation of strings.

In conclusion we want to thank 
the CICYT and the Xunta de Galicia  for financial
support. We strongly thank to N. Armesto and M. A. Braun who participate in
the earliest stage of this work.

\newpage
\noindent{\Large {\bf Figure Captions}}

\vskip 0.5cm

\noindent{\bf Figure 1.} Rapidity distributions of $h^-$ (a),
$K^0_s$ (b) and $\overline{\Lambda}$ (c) 
obtained with the SFMC code with string fusion and compared to experimental
data (black squares) \cite{NA49,NA49b} 
for Pb--Pb central ($b \leq 2.8$ fm) collisions at $p_{lab}=158$
AGeV/c.

\vskip 0.5cm

\noindent{\bf Figure 2.} Rapidity distribution (a) 
and $|x_F|$ distribution (b) (taking into account both
negative $x_F$ and positive $x_F$ contributions) 
of antiprotons around the
kinematical limits for Pb--Pb collisions at 158 AGeV/c. The continuous line
corresponds to the SFMC results with string fusion mechanism, the dashed line
is the SFMC result without string fusion
and the black squares are the experimental data points
\cite{NA52}.

\vskip 0.5cm

\noindent{\bf Figure 3.} $x_F$ distributions for $x_F$ $>$ 1 in Pb--Pb
central collisions at $p_{lab}=200$ AGeV/c of
protons (a) and $\Lambda$ (b)  with (continuous line) and without
(dashed line)
string fusion.

%\end{document}

%%%%%%%%%%%%%%%%%%%%%%%%%%%%
\textwidth    170mm
\textheight   300mm
\columnsep     38pt
\topmargin    -1.3 cm
\oddsidemargin  -0.9pt
%%%%%%%%%%%%%%%%%%%%%%%%%%%%

%\pagestyle{empty}

\newpage
\centerline{\bf Fig. 1}
\begin{figure}[hbtp]
\begin{center}
\mbox{\epsfig{file=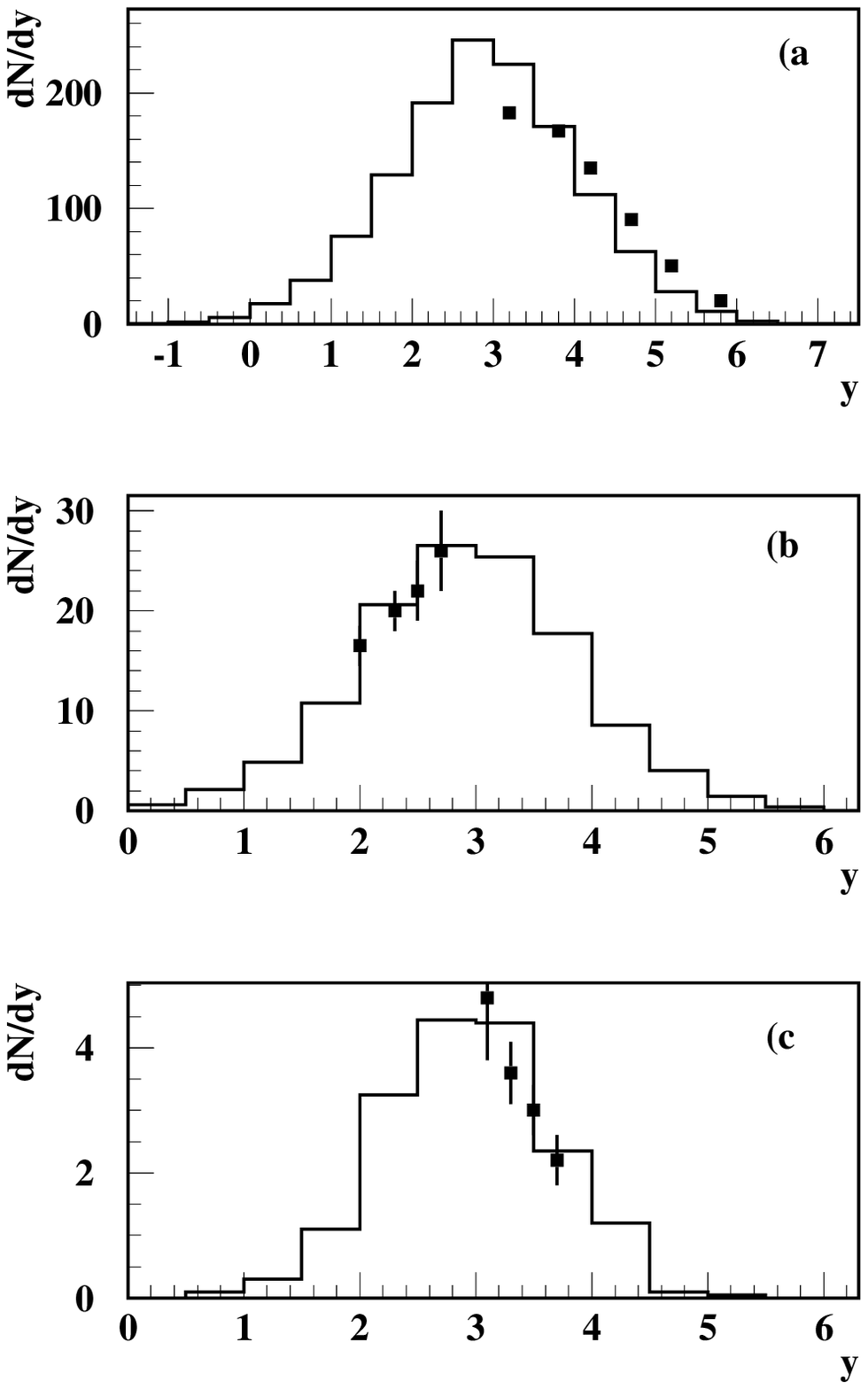,height=18cm}}
\end{center}
\end{figure}

\pagestyle{empty}

\newpage
\centerline{\bf Fig. 2}
\begin{figure}[hbtp]
\begin{center}
\mbox{\epsfig{file=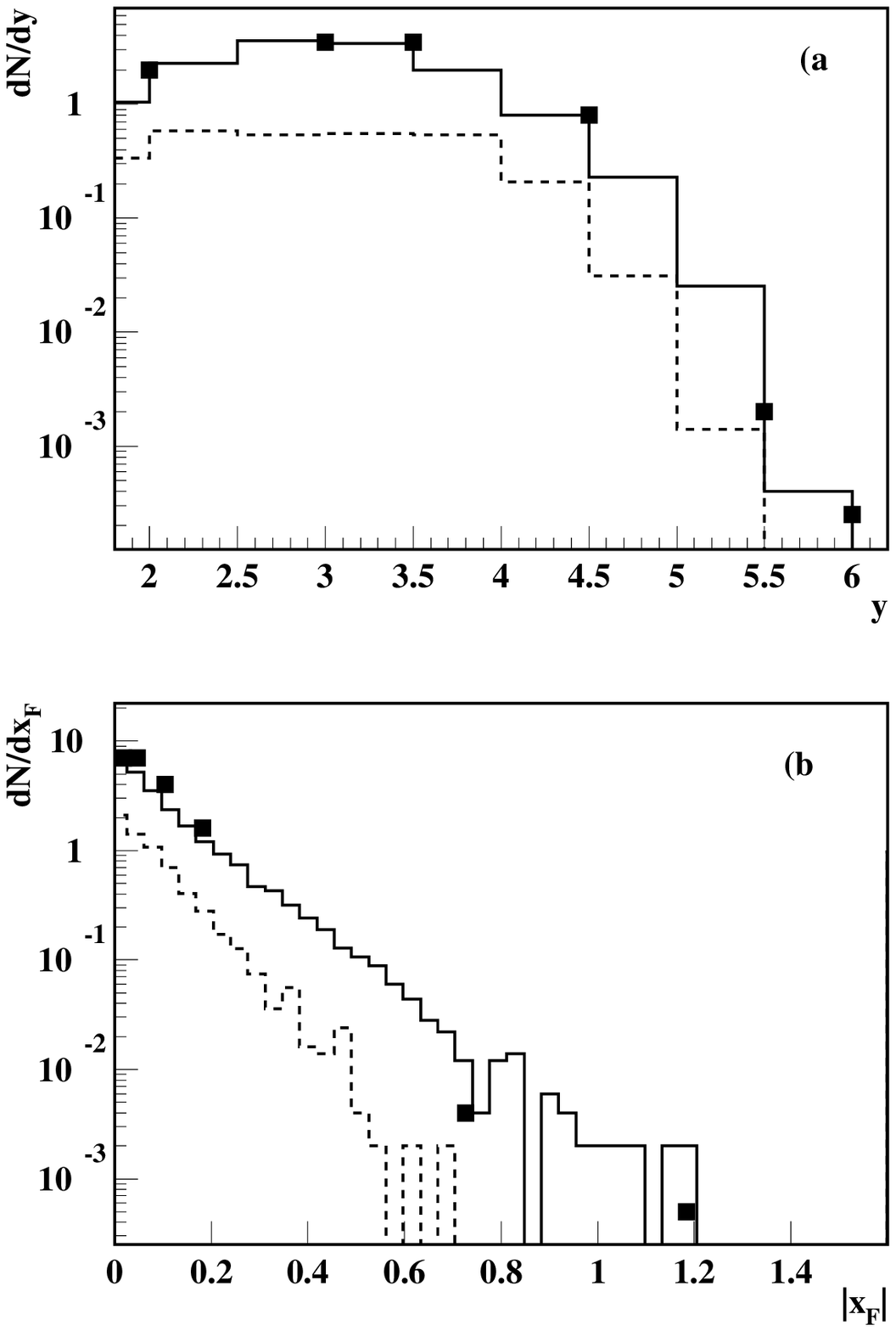,height=17cm}}
\end{center}
\end{figure}

\newpage
\centerline{\bf Fig. 3}
\begin{figure}[hbtp]
\begin{center}
\mbox{\epsfig{file=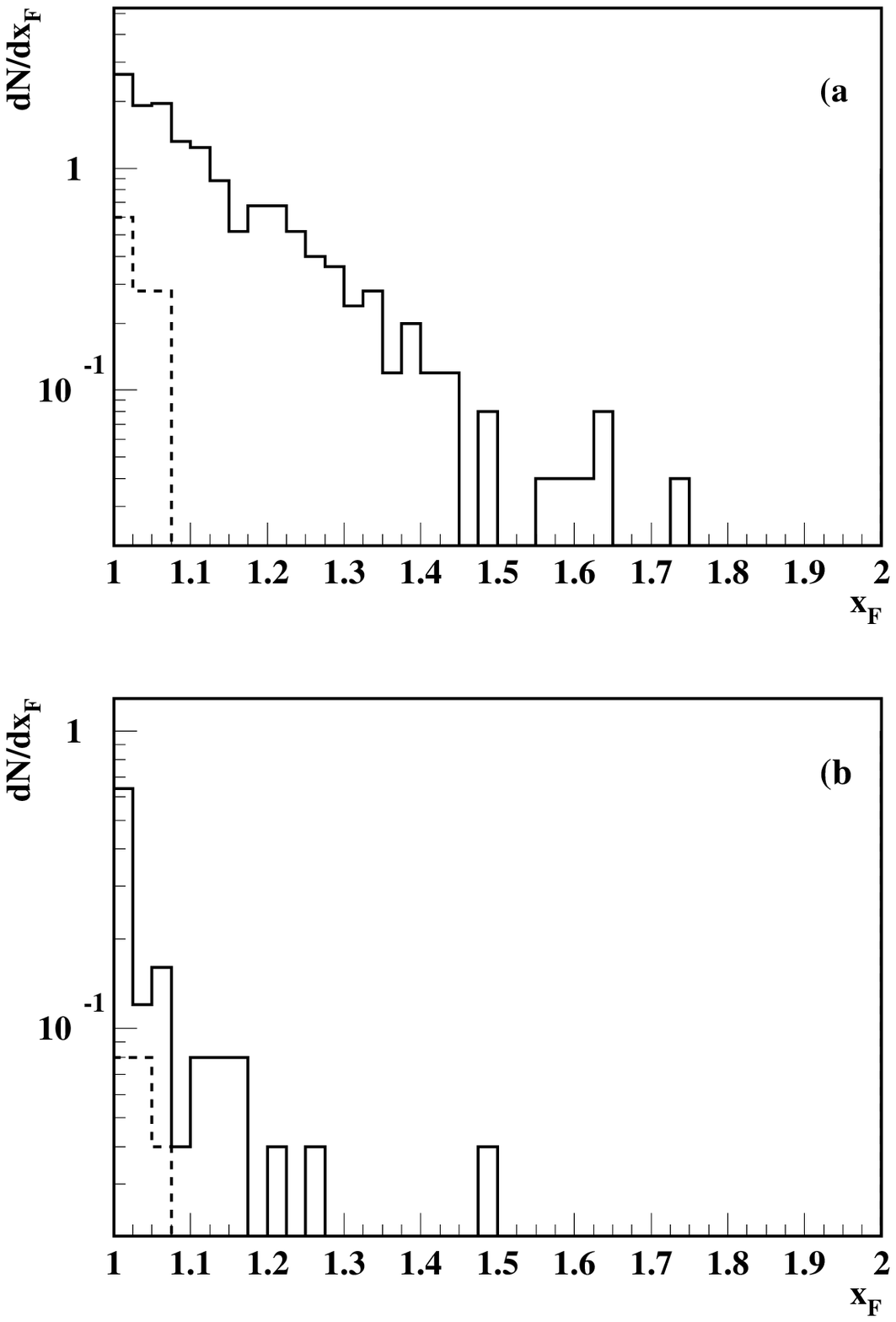,height=17cm}}
\end{center}
\end{figure}


\begin{thebibliography}{99}           

%%%%%%%%%%%%%%%%

\bibitem{ee} Th. Ullrich, CERES/NA45 Collaboration, Nucl. Phys. {\bf A610},
317c (1996).

\bibitem{str} H. Helstrup, WA97 Collaboration, Nucl. Phys. {\bf A610},
165c (1996).

\bibitem{supre} M. Gonin, NA50 Collaboration, Nucl. Phys. {\bf A610},
404c (1996).

\bibitem{QGPform} C. M. Ko, G. Q. Li, G. E. Brown and H. Sorge, Nucl. Phys.
{\bf A610}, 342c (1996); G. Q. Li, C. M. Ko and G. E. Brown, Phys. Rev. Lett.
{\bf 75}, 4007 (1995); W. Cassing, W. Ehehalt and C. M. Ko, Phys. Lett. {\bf
B363}, 35 (1995).

\bibitem{Bla96} J. P. Blaizot and J. Y. Ollitrault, 
Phys. Rev. Lett. {\bf 77}, 1703 (1996).

\bibitem{Kha96}
D. Kharzeev and H. Satz, Phys. Lett. {\bf B336}, 316 (1996).

\bibitem{Won96} C.--Y. Wong, Nucl. Phys. {\bf A610},
434c (1996).

\bibitem{Perco} N. Armesto, M. A. Braun, E. G. Ferreiro and C. Pajares, Phys.
Rev. Lett. {\bf 77}, 3736 (1996).
                                       
\bibitem{Gav96} S. Gavin and R. Vogt, Phys.
Rev. Lett. {\bf 78}, 1006 (1997).

\bibitem{Comover} A. Capella, A. Kaidalov, A. Kouider Akil and C. Gerschel,
Phys. Lett. {\bf B393}, 431 (1997); 
N. Armesto and A. Capella, preprint LPTHE Orsay 97/11 (1997).

\bibitem{Sorge} H. Sorge, E. Shuryak and I. Zahed, 
preprint hep--ph/9705329 (1997).


\bibitem{NA52} R. Klingerberg, NA52 Collaboration, Nucl. Phys. {\bf A610}
306c, (1996); S. Kabana, NA52 Collaboration, Proceedings of the International
Symposium on Strangeness and Quark Matter, Santorini, Grecee, April 1997. To
appear in Journal of Physics G: Nuclear and Particle
Physics.

%%%%%%%%%%%%%%%%%%%%%%%%%%%%%%%%%%%%%%%%%%%%%%%%%%%%%%%%%
\bibitem{Bal74} A. M. Baldin {\it et al.}, Yad. Fiz. {\bf 20}, 1210 (1974);
Sov.
 J. Nucl. Phys. {\bf 20}, 629 (1975).

\bibitem{Str81} M. I. Strikman and L. L. Frankfurt, Phys. Rep. {\bf 76}, 215
(1981).

\bibitem{Efr88} A. V. Efremov, A. B. Kaidalov,
V. T. Kim, G. I. Lykasov and
N. V. Slavin, Sov. J. Nucl. Phys. {\bf 47}, 868 (1988).

\bibitem{Bra94} M. A. Braun and V. Vechermin, Nucl. Phys. {\bf B427}, 614
(1994).

\bibitem{xfey}  N. Armesto,
M. A. Braun, E. G. Ferreiro, C. Pajares and Yu. M. Shabelski, Phys. Lett. {\bf
B389}, 78 (1996); Astroparticle Physics {\bf 6}, 329 (1997).

\bibitem{SFMC} N. S. Amelin, M. A. Braun and C. Pajares,
Phys. Lett. {\bf B306}, 312 (1993); Z. Phys. {\bf C63}, 507 (1994).

\bibitem{strange} N. Armesto, M. A. Braun, E. G. Ferreiro and C. Pajares,
Phys. Lett. {\bf B344} (1995) 301.

%%%%%%%%%%%%%%%%%%%%%%%%%%%%%%%%%%%%%%%%%%%%%%%%%%%%%%%%%%%%%

\bibitem{Cap94} A. Capella, U. P. Sukhatme, C.--I. Tan and J.
Tran Thanh Van, Phys. Rep. {\bf 236}, 225 (1994).

\bibitem{Kai84} A. B. Kaidalov and K. A. Ter--Martirosyan, Phys. Lett. {\bf
B117}, 247 (1982).

\bibitem{Wer87} K. Werner, Phys. Rep. {\bf 232}, 87 (1993).

\bibitem{NA49} P. G. Jones, NA49 Collaboration, Nucl. Phys. {\bf A610}, 
188c (1996).

\bibitem{NA49b} C. Borman, NA49 Collaboration, Proceedings of the International
Symposium on Strangeness and Quark Matter, Santorini, Grecee, April 1997. To
appear in Journal of Physics G: Nuclear and Particle
Physics.

\bibitem{SQM97} E. G. Ferreiro and C. Pajares, Proceedings of the International
Symposium on Strangeness and Quark Matter, Santorini, Grecee, April 1997. To
appear in Journal of Physics G: Nuclear and Particle
Physics.

\bibitem{Bo} B. Andersson and P. Henning, Nucl. Phys. {\bf B355}, 82 (1991).

\end{thebibliography}
\end{document}